\documentclass[sigplan]{acmart}
\pdfoutput=1
\date{}
\usepackage{algorithm}
\usepackage{amsmath,array,color,colortbl,graphicx,multirow} %%%%chen%%%%%
\usepackage{comment}
\usepackage{balance}
\usepackage{xspace}
\usepackage{enumitem}
\usepackage{makecell}
\usepackage{wrapfig}

\newtheorem{theorem}{Theorem}
\newtheorem{observation}{Observation}

\def\Expct{\mathbb{E}}

\def\M{\mathcal{M}}
\def\D{\mathcal{D}}
\def\dct{\mathrm{DCT}}

\definecolor{applegreen}{rgb}{0.55, 0.71, 0.0}

\newcommand{\system}{{\sc{Hybrid}}\xspace}

\def\uni{U}
\def\skd{S}

\newcommand{\sflow}{s}
\newcommand{\mflow}{m}
\newcommand{\lflow}{\ell}

\newcommand{\snum}{k_{s}}
\newcommand{\rnum}{k_{r}}
\newcommand{\cnum}{k_{c}}

\newcommand{\sys}{sys}

\newcommand{\EPL}{\mathrm{epl}}

\newcommand{\rrec}{R_{r}}
\newcommand{\crec}{R_{c}}

\date{}

\usepackage{blindtext}
\usepackage{color}
\usepackage{url}
\usepackage{xspace}
\usepackage{amsthm}

\usepackage{mathtools}
\usepackage{float}
\usepackage{graphicx}
\usepackage{footnote}
\usepackage{comment}
\makesavenoteenv{tabular}
\makesavenoteenv{table}
\usepackage{multirow}
\usepackage{booktabs}
\usepackage[flushleft]{threeparttable}
\usepackage{tikz}
\usetikzlibrary{calc}
\usepackage{pgfplots}
    \usepgfplotslibrary{colorbrewer}
    \pgfplotsset{
        cycle list/Set1-9,
        cycle multiindex* list={
            mark list*\nextlist
            Set1-9\nextlist
        },
    }
    \usepgfplotslibrary{fillbetween}
    \usetikzlibrary{patterns}
\usepackage{algpseudocode}
\usepackage[acronym]{glossaries}
\usepackage[utf8]{inputenc}
\usepackage[english]{babel}

\renewcommand\footnotetextcopyrightpermission[1]{}
\usepackage[font=small]{caption}

\newcommand{\para}[1]{\noindent \textbf{#1}\xspace}
 \renewcommand\footnotetextcopyrightpermission[1]{} % removes footnote with conference info
 \setcopyright{none}
 \acmDOI{}
 \acmISBN{}
 \acmPrice{}
\newcommand{\rot}{RotorNet\xspace}

\def\card#1{\lvert #1 \rvert}

\usepackage{amsfonts}

\renewcommand{\ss}{\textsc{Static}\xspace}
\newcommand{\rs}{\textsc{Rotor}\xspace}
\newcommand{\cs}{\textsc{Demand-aware}\xspace}
\newcommand{\ssn}{\emph{static-net}\xspace}
\newcommand{\rsn}{\emph{rotor-net}\xspace}
\newcommand{\esn}{\emph{expander-net}\xspace}

\newcommand{\csn}{\emph{demand-aware-net}\xspace}

\newcommand{\longPL}[1]{\ignorespaces}

\newcommand{\model}{\textsc{TMT}\xspace}
\newcommand{\s}{\mathcal{S}\xspace}
\renewcommand{\r}{\mathcal{R}\xspace}
\renewcommand{\c}{\mathcal{C}\xspace}

\def\card#1{\lvert #1 \rvert}
  
\def\compactify{\itemsep=0pt \topsep=0pt \partopsep=0pt \parsep=0pt}
\let\latexusecounter=\usecounter

\usepackage{color}
\definecolor{lightgray}{gray}{0.75}

\usepackage[noframe]{showframe}
\usepackage{framed}
\usepackage{lipsum}
 {\endMakeFramed}
\definecolor{shadecolor}{gray}{0.75}

\DeclareMathOperator*{\argmax}{arg\,max}

\begin{document}

\title{Performance Analysis of Demand-Oblivious and Demand-Aware Optical Datacenter Network Designs}

\author{Chen Griner$^1$ \quad Johannes Zerwas$^2$ \quad Andreas Blenk$^2$ \\ Manya Ghobadi$^3$ \quad Stefan Schmid$^4$
\quad Chen Avin$^1$\\
{\small $^1$ BGU, Israel \quad $^2$ TU Munich, Germany \quad $^3$ MIT, USA \quad $^4$ University of Vienna, Austria } 
}

\sloppy

\begin{abstract}
This paper presents a performance analysis of 
the design space of optical datacenter networks,
including both demand-oblivious (static or dynamic)
and demand-aware networks. 
We formally show that the
number of specific optical switch
types which should be used in an optimized
datacenter network, depends on the traffic pattern,
and in particular, the flow size distribution. 
\end{abstract}

\maketitle

\sloppy

\section{Introduction}
\label{sec:intro}

This paper is motivated by the observation that
different traffic patterns (and in particular: different flow
size distributions), require different datacenter network designs
(and in particular: different distributions of optical switch types).
We argue that a mismatch 
between some common traffic patterns and the optical switching technologies used to serve them can be quantified in terms of \emph{taxes}.

Flow size distributions, and in turn flow transmission times, can vary widely across and within applications. %This is important as  flow size determines the flow transmission time. 
For example, the ideal flow transmission time on a $40Gbps$ link 
can range from micro-seconds for small flows up to seconds for elephant flows.
The flow transmission time however does not only depend on the flow size and the link speed, but also on the used optical switching technology and the reconfiguration time: 
whereas there are no reconfigurations in static topologies, dynamic topologies entail a reconfiguration time. For example, 
the reconfiguration time of rotor switches can be in the order of $10\mu s$ per slot \cite{rotornet}  
whereas the reconfiguration time of demand-aware optical switches may be in the order of $15ms$ \cite{MEMSMatrixOpticalSwitches, calient}.
While the exact reconfiguration times will depend on the specific technology, the reconfiguration times of demand-aware topologies are likely to be higher than those of demand-oblivious topologies: demand-awareness may require data collection and running optimization algorithms. 
 
A first key observation is that whether a flow can profit from reconfigurations depends on its size. For example, for an elephant flow whose transmission time is large compared to the reconfiguration time, the reconfiguration can be amortized and may pay off in the long run: while a demand-aware dynamic topology first needs to be configured (which takes a fixed amount of time), it then allows to provide direct connectivity to elephant flows.
In contrast, for a mice flow, the reconfiguration time may be significantly higher than the flow completion time in a static network.
In general, we can regard the reconfiguration time as 
a ``latency tax'' (or ``reconfiguration tax'').
%take away
A latency
tax can be amortized by the long transmission times of large flows.

While static topologies do not introduce any latency tax, they require multi-hop forwarding. This is problematic especially for large flows: the more hops a flow has to traverse, the more network capacity is consumed. This can be seen as a ``bandwidth tax'' (or an ``infrastructure tax''), as noticed in prior work~\cite{rotornet}.
Regarding the design choice between static vs dynamic topology
%%Figure
%(cf. Figure \ref{fig:types}), 
%End %Figure
we therefore observe:
Whereas dynamic topologies
introduce a latency tax, static topologies 
introduce a bandwidth tax. 

Dynamic topologies can reduce the bandwidth tax by avoiding multi-hop forwarding. For instance, rotor switches provide periodic \emph{direct} connectivity, and have been shown to perform particularly well for uniform all-to-all traffic patterns~\cite{rotornet,opera}. However, such demand-oblivious dynamic architectures are not optimal for the elephant flows;
as typically a majority of the bytes are carried by elephant flows, optimizing for these large flows is important. % (see e.g., our Datamining workload). 
While in principle, 
Valiant routing \cite{valiant1982scheme} can be used in combination with rotor switches
to carry large flows, 
this again results in bandwidth tax.
In contrast, in a demand-aware topology, directed shortcuts can be set up specifically for such elephant flows.
Regarding the design choice between demand-oblivious vs demand-aware topology,
we observe that whereas demand-oblivious topologies perform well under uniform demands, demand-aware topologies perform well under skewed traffic and large flows
for which latency taxes can be amortized.

This paper provides a mathematical model that can decide on the optimal flow assignments to each topology type based on the flow size distributions. 

\section{Topology and Switch Models}
\label{sec:design}

Motivated by our observations above,
%the need for a unified network design by highlighting the observation that static, rotor, and %demand-aware switches come with different advantages and disadvantages. 
%Accordingly, in the next sections 
we consider a unified datacenter architecture
which combines all three switch types:
dynamic oblivious (rotor), dynamic demand-aware.
and static.
We first present our datacenter network model
and then model the different possible topology components.
These models, 
will allow us to compute the optimal size and composition of
different topology types,
to improve performance,
depending on link rates and reconfiguration times
as well as the given traffic mix.

\subsection{Topology Models}

Our datacenter network model relies on 
a two-layer leaf-spine network architecture with $n$ leaf switches that are ToR switches and $k$ optical
spine switches that can be of different types: 
\ss, \rs and \cs; we will denote the number of 
switches from each type 
by $\snum, \rnum$ and $\cnum$ respectively where $k= \snum + \rnum + \cnum$.
We will discuss how to compute the value for each type later.
Since each optical spine switch connects its in-out port via a \emph{matching}
we will refer to it as the ToR-Matching-ToR (\model) network model.
This network architecture generalizes existing
architectures such as RotorNet \cite{rotornet}, Opera \cite{opera}
and Sirius \cite{ballani2020sirius}, by supporting multiple switch types.

More specifically, the \model  network 
interconnects a set $N$ of $n$ ToRs, $\{1,2, \dots, n\}$
and its two-layer leaf-spine architecture composed of
 leaf switches and spine switches. %, similar  to~\cite{opera,rotornet}. 
The $n$ ToR packet switches are connected using $k$ spine  switches, 
$SW = \{sw_1, sw_2, \dots, sw_k\}$ and each switch internally connects its 
in-out ports via a matching. 
We assume that each ToR $i: 1 \le i \le n$ has $k$ uplinks, where uplink $j: 1 \le j \le k$  connects to port $i$ in $sw_j$.  The directed outgoing (leaf) uplink is connected to the incoming port of the (spine) switch, and the directed incoming (leaf) uplink is connected to the outgoing port of the (spine) switch. Each spine switch has $n$ input ports and $n$ output ports and the connections are directed, from input to output ports.
As a running example,
throughout this paper we assume that the rate of each link is $r=40Gbps$; however our model is general and can be used for any bandwidth as we discuss later: similar results to what we report here hold also for lower rates (e.g., 10Gbps as in \cite{opera,rotornet,xpander}) and higher rates, e.g. 100Gbps.

At any point in time, each switch  $sw \in SW$ provides a \emph{matching} between its input  and output ports. Depending on the switch type, this matching may be \emph{reconfigured} at runtime to another matching. Each switch $j$ has a set of matchings $\M_j$ of size $m_j=\card{\mathcal{M}_j}$ and $m_j$ may be larger than one. %, i.e., $m_j=\card{\mathcal{M}_j}>1$. 
Changing from a matching $M' \in \mathcal{M}$ to a matching $M''\in \mathcal{M}$ takes time, which we model with a parameter $R_j$:  the \emph{reconfiguration time} of switch $j$. During reconfiguration, the links in $M' \cap M''$, i.e., the links which are not being reconfigured, can still be used for forwarding; the remaining links are blocked during the reconfiguration \cite{huang2016sunflow}. Depending on the technology, different switches in $SW$ support different sets of matchings and reconfiguration times.

\subsection{Switch Models}\label{ssec:compos}

Our network model can be instantiated with different switches, accounting for the different and specific switch characteristics, thereby by creating different typology types or components.  In this paper, we consider  three fundamental topology components: a static part, a rotor-only part, and a demand-aware part. These components may either be realized by different switch technologies, or by a single switch technology that can support multiple \text{modes of operation}. The former may be more cost effective (e.g., static topologies are cheaper), while the latter is more flexible. 

Abstractly, the three topologies can be described in a 
unified form using a collection of spine switches. Each spine switch type, in turn is defined by a 4-tuple,
$sw=(m, \M, S, R)$
%\noindent
where  $m$  is the number of matchings the switch can
support;
$\M$ is the specific set or sequence of $m$ matchings 
the switch can realize;
$S$ is the minimal circuit-hold time a switch needs to remain
in a specific matching before switching to the next matching;
$R$ is the reconfiguration time of changing between matchings.
Using our notation, the three topologies can be formalized as follows:

%\item 
\para{\textbf{Demand-aware topology (\cs):}} 
We create a demand-aware topology using a collection of
$\cnum$ \cs reconfigurable switches. Each such switch is described by the tuple
$sw=(n!, \mathcal{M}, S, \crec)$.
Demand-aware switches have the freedom to flexibly
reconfigure to \emph{any} of the $m=n!$ 
possible matchings; i.e., $\mathcal{M}= S_n$ where $S_n$ is the symmetric group of $[1...n]$.
%\stefan{set notation for permutation}. 
The demand-aware switch can be implemented using off-the-shelf 3D MEMS 
technology with reconfiguration time in the order of tens of $ms$. 
In this paper, we assume $\crec=15ms$ which is the typical reconfiguration time of a 3D MEMS switch \cite{calient}.
The circuit-hold time $S$ can change during the operation of the \cs switch, 
but as a rule of thumb $S \gg \crec$ for the reconfiguration to be worthwhile.

%\item 
\para{\textbf{Rotor-based topology (\rs):}}
A rotor-based topology consists of the union of $\rnum$ 
\rs switches. Each  switch is described by the tuple $sw=(n-1,\mathcal{M}, \delta, \rrec)$:
a \rs switch cycles through  $n-1$ matchings 
specified by $\mathcal{M}$, emulating a fully-connected
network (i.e., complete graph) and hence providing high bandwidth to all-to-all traffic. 
Our \rs switch is a slight generalization of the original \rs switches~\cite{rotornet}
used in \rot since our model uses $n-1$ matchings and not $n/k$ matchings as proposed originally.
%(We will explain why this generalization improves the performance of \rot in the technical report.) 
The reconfiguration time for the \rs switch is in the order of few $\mu s$; here we assume $\rrec=10 \mu s$ as in \cite{opera,rotornet}.
The circuit-hold time of a \rs switch is called the \emph{slot} time and is denoted as $\delta$.  The slot time is tunable and depends on the reconfiguration time, 
where a reasonable setup is at least $\delta=9\rrec = 90 \mu s$ to reach 90\% amortization of the reconfiguration time  \cite{opera,rotornet}.
%\stefan{Chen: explain} \chen{will add to appendix. not here},
%\andi{If the links are not bidirectional, this would not be the complete graph, or not?} \chen{It will be}

%\item 
\para{\textbf{Static topology (\ss):}}
We describe the static topology as a union
of $\snum$ matchings, where each matching can be implemented, e.g., using an
optical patch panel (our analysis also applies to electrical static topologies).
In the case of a static component, the 4-tuple switch specification can be represented by:  $sw=(1,\mathcal{M}, \infty, 0)$ 
where the component provides a single (i.e., $m=1$) predefined matching $\{M\}=\mathcal{M}$ that does not change over time ($S=\infty, R=0$).
%\footnote{We call this a switch for ease of presentation. In practice, once configured with a matching it has static forwarding rules.}.
%We can use a combination of these switches to create different static network topologies. %, such as fat-trees~\cite{fat-tree}. \chen{really?}
The static switches are cost-effective components to create regular graphs, such as expander graphs, providing low latency
for short flows using multi-hop routing. 
%Prior work has shown that
Good expanders can be obtained by taking the
union of a few matchings~\cite{goldreich2011basic}.

While the \ss~and \rs-based topologies 
are based on demand-oblivious switches, a
\cs topology is adapting to the demand. 
%reconfiguration decisions in \cs are more complex and 
Demand-aware switches are hence likely more complex for their support of greater matching flexibility,
and their reconfiguration times as mentioned above are naturally higher, i.e., $\rrec \ll \crec$. Moreover, a
demand-aware topology requires a control logic to decide on which on-demand links to establish based on the current traffic demand~\cite{cthrough, helios, reactor}.

Since the TMT network can be configured
with multiple different switch types and hence
contains multiple types of topology components,
we introduce the following terminology.
Let $(\snum,\rnum,\cnum)$ denote a network consisting of
$\snum$ static switches, $\rnum$ rotor switches,
and $\cnum$ demand-aware switches.
We will refer to a network consisting only of  $k$ 
\ss switches, i.e., a network with $(k,0,0)$, as \ssn; 
as we will see, we will assume that the static topology component of 
relies on expander graphs, and
we will hence refer to this network as \esn.
We will further refer to the network consisting of only 
\rs switches, i.e., networks with $(0,k,0)$, as \rsn,
and to topology components consisting only of
\cs switches, i.e., a network with $(0,0,k)$, as \csn.

We also note that the TMT network can be used to model
many existing systems. For example,
RotorNet \cite{rotornet}, 
Opera \cite{opera} and Sirius \cite{sirius}
rely on periodic matchings and
can be modelled as a \rsn.  
Networks like ProjecToR \cite{projector},
Eclipse \cite{venkatakrishnan2018costly},
Helios \cite{helios}, ReNet \cite{apocs21renets},
BMA \cite{perf20bmatch}, among others \cite{ancs18}, rely on demand-aware
matchings. To be more specific, while for example
the demand-aware links of ProjecToR 
are based on free space optics, conceptually
it can still be modelled as a \csn; 
ProjecToR additionally uses a 
static electric network, which in our conceptual model can
also be described using an \esn.
Our model and analysis also applies to Xpander~\cite{xpander}, which 
can be modelled as an \esn as well (even though it is based on electrical switches).
We refer the reader to the related work section for additional details.

%\chen{need a sentence what next....}

\section{Performance Analysis} 
%\section{Throughput and Demand Completion Time Analysis} 
%: Analytical Study} %Architecture and Analysis}
\label{sec:brain}

This section presents an analysis of 
$(i)$ the optimal topology parameters for each type of topology; $(ii)$ the flow assignment strategy for each topology; and ($iii$) the throughput and flow completion times for the system as a whole. To do this we first describe the traffic generation model we consider.

\subsection{Flow Assignment Model} 

Throughout this paper, we consider the following flow assignment algorithm:
given a $(\snum,\rnum,\cnum)$ network, we divide flows into three categories, based on  
three sizes: \emph{small} ($\sflow$), \emph{medium} ($\mflow$)  
and \emph{large} ($\lflow$) flows.
The size \emph{thresholds} to assign flows to these categories are denoted by $\card{\mflow}$ and $\card{\lflow}$. Namely, \emph{small} flows are of size less than $\card{\mflow}$, \emph{medium} flows are of size more than  $\card{\mflow}$ and less than $\card{\lflow}$, and \emph{large} flows are of size larger than $\card{\lflow}$. 

We first determine the size threshold for medium flows, $\card{\mflow}$, as the slot time in \rs switches, i.e., $\card{\mflow}=\delta r$. This is done to insure low delay for small flows and good link utilization for medium flows; we then set the large flow size threshold, $\card{\lflow}$, as the minimum flow size that will have shorter completion time on \cs switches, see Eq. \eqref{eq:threshold} in the next section. 

\begin{algorithm}[t]
\caption{Flow assignment}\label{alg:cachenet}
	\begin{algorithmic}[1]
%		\Require something
%		\Ensure something
		\State \textbf{Switch} depending on flow size
		\State  ~~~~ \textbf{Case} small flow:          \hfill \Comment{latency-sensitive flow}
			\State~~~~~~~~ send to \emph{static expander}                 \hfill \Comment{using multi-hop}
		\State ~~~~ \textbf{Case} large flow:
		\State	~~~~~~~~ \textbf{If} a direct link is available to reconfigure: 
		\State	~~~~~~~~~~	send to \emph{demand-aware} topology                \hfill \Comment{single hop}
			\State ~~~~~~~~  \textbf{Else}                                  \hfill \Comment{Under provisioned demand-aware}
			\State	~~~~~~~~~~	send to \emph{rotor-based} topology             \hfill \Comment{using 1 or 2 hops} 	 
		\State  ~~~~ \textbf{Case} medium flow:
%		\State	~~~~~~~~ \textbf{If} a direct link is available to reconfigure:
		\State	~~~~~~~~~~ send to \emph{rotor-based} topology   \label{alg:mid}          \hfill \Comment{using 1 or 2 hops}
%		\State ~~~~~~~~  \textbf{Else}
%		\State	~~~~~~~~~~ send to \rsn (using 1 or 2 hops)
	\end{algorithmic}
\end{algorithm}

Algorithm~\ref{alg:cachenet} 
describes how we distribute the 
traffic classes among the three switch types: 
small, latency-sensitive flows are 
forwarded via a static expander built from 
\ss switches; 
large flows are transmitted
via the $\cnum$ many \cs switches in the system;
and the remaining (medium) flows
describing e.g., all-to-all traffic which is not 
latency-sensitive, are routed via the $\rnum$ \rs switches. 
We can manage the large flows using an approach
which can be seen as a distributed link cache \cite{perf20bmatch}:
when a new demand-aware connection needs to be established,
an existing link must be replaced or ``evicted''.
%The demand-aware matching is hence computed in a decentralized
%manner, and can be configured with different cache replacement policies \cite{smith1982cache} and/or matrix decompositions %\cite{venkatakrishnan2018costly}.      
While this introduces interesting optimization opportunities,
in the following, we will focus on a simple strategy:
when a large flow should be sent to the \cs switches, but there are no available ports to serve it (the related source/destination ports are already serving other flows), we greedily send the large flow to be served by \rs switches. 
When this happens continuously, we say that the demand-aware switches are \emph{under provisioned}. In the next section, we derive $\cnum^*$, the optimal number of \cs switches (under a given traffic assumption) which  minimizes forwarding of large flows over  
\rs switches.

\subsection{Traffic Generation Model and Metrics}

Inspired by prior work~\cite{opera,xpander,sirius}, we consider the following traffic model:
flows arrive over time, according to a sequence $\sigma=
(f_1,f_2,\ldots)$ where the $f_i$s
are individual flows. Each flow $f_i=(s_i, d_i, \Delta_i)$ has a source rack $s_i \in N$, a destination  rack $d_i \in N$, and a size $f_i$ 
(in bytes)\footnote{Flows are originally generated by servers, 
but our focus here is on the flows from a rack-level 
perspective.}. 
For our analysis we study two fundamental models 
to generate a load $0 \le x \le 1$:
%\begin{itemize}[leftmargin=1em]
	%\item 

\para{\textbf{Uniform Traffic, $\uni(x)$}:} 
	In this model, each ToR serves  
traffic at an average rate of $x \cdot k \cdot r$, where $r$ is the line rate and $k$ the number of uplinks (and spine switches).
Flow arrival times follow a Poisson process,
flow sizes are sampled from a distribution $\D$,
and for each source ToR the destination ToR is
chosen uniformly at random from $N$. 
The expected total amount of traffic 
generated per second is $n \cdot \uni(x) = n k x r$ (recall then $n$ is  the number of ToRs).

%\item 
\para{\textbf{Skewed Traffic, $\skd(x)$:}} Inspired by recent datacenter studies~\cite{projector, roy2015inside, fat-free}, in this model only a random fraction of 
$x$ ToRs 
are active, but 
each active ToR operates   at
100\% throughput, i.e., the $k$  ports are fully utilized,
at line rate, $r$. More generally $\skd_L(x)$ means that each active ToR operates at throughput $L\cdot 100\%$ and  $\skd(x)= \skd_1(x)$.
Again, flow arrivals follow a Poisson
process, flow sizes are sampled 
from $\D$, and destinations are chosen uniformly at random from the set of \emph{active} ToRs. %\manya{servers?}. 
The expected total amount of traffic that the system generates and needs to serve
per second is $x \cdot n \cdot \uni(1) = x n k r$, i.e., each active ToR serves traffic of $\uni(1)$.
%\end{description}
%\end{itemize}

We study analytically the \emph{demand completion time} ($\dct$)
as the metric of interest:
the total time it
 takes to serve an accumulated demand matrix built
from a collection of flows that arrived in one second, 
%assuming that all flows arrived at time zero, 
either according to the $\uni(x)$ or $\skd(x)$ traffic generation models.
The demand completion time is a measure for the \emph{capacity} of the network:
if for a given $x$ the completion time is less than $1 s$, this means
that the network has sufficient capacity to serve this load. 
In the uniform traffic model, for a given load (throughput) $x$, we are interested in the demand completion time of $\dct(\uni(x))$.
In the skewed traffic model, for a given fraction of active ToRs $x$, we are interested in the maximum throughput $L$, s.t. $\dct(\skd_L(x)) \le 1$; in particular, we are interested whether the network is 
\emph{throughput-proportional}.
Following the definitions and observation in~\cite{xpander} 
%Figure
%(see also Figure~2 in~\cite{xpander})
%End Figure
, a network is  \emph{throughput-proportional} when it is ``able to distribute its capacity evenly across the set of servers with actual traffic demands''~\cite{xpander}.
The \emph{non-active} ToRs can share their capacity and help the $x$ fraction of \emph{active} ToRs. Hence, $L$ is proportional to the total generated load and network capacity, and not only to $x$. 

%\chen{add throughput-proportional definition here?}
\subsection{Analysis of Parameters and Completion Times}
%\subsection{Analysis of the Topology Parameters and Demand Completion Times}
In this section, we present our theoretical results on the demand completion times 
of \esn, \rsn and our combined model, henceforth called \system, in both the $\uni(x)$ and the $\skd(x)$ traffic models.

We start by considering the uniform traffic model $\uni(x)$. 
We show that for all considered systems, 
\esn, \rsn and \system, the demand completion time grows (almost) linearly with load; 
%\manya{isn't this obvious? is there anything complicated/surprising here?}
moreover, for a wide range of parameters (as we will show later), \system has consistently lower demand completion times 
and therefore higher effective capacity of the system.

Formally, for a demand distribution $D$ and using $k$ switches, let $\dct(\sys, D, k)$ denote the demand completion time of a system 
$\sys\in\{\esn,\rsn,\system\}$.  Recall that $\esn = (k,0,0), \rsn = (0, k, 0)$ and 
$\system=(\snum, \rnum, \cnum)$.
Let $\tau$ denote the type of a flow where 
$\tau \in \{\sflow, \mflow, \lflow\}$
%is the flow type, 
and let $\uni(x, \tau)$ 
%\#{\tau}$ 
denote the expected number of bytes per second in flows of type $\tau$ 
when a ToR is working at a fraction $x$ of its 100\% rate, that is, 
when all $k$ uplinks are sending traffic, each at rate $x r ~Gbps$, toward the spine switches.
Note that $\uni(1, \sflow) + \uni(1, \mflow) + \uni(1, \lflow) = kr$.
Let $\phi$ denote the \emph{traffic skewness} of the flow size distribution. Formally, $\phi$ denotes the fraction of packets (or bytes) sent in \rsn via a single hop, and $1-\phi$ is the fraction of packets that are sent using 
Valiant routing~\cite{rotornet}, using two hops.  
We approximate $\phi$ for a given distribution and the load (or empirical distribution) as one minus the variation distance~\cite{rosenthal1995convergence} from the uniform distribution.
This means that links (or source-destination pairs) that are active above the average load can send packets via links that are below the average load (using two hops).
%We discuss $\phi$ in detail in the technical report.
%in Appendix \ref{sec:flowanalysis}.  % \khen{so now it is in an other section}

We can state the following formal results about the demand completion time of the different systems:

\begin{theorem}[Uniform Traffic]\label{thm:linear}
Consider the uniform traffic model, $\uni(x)$ with 
a flow size distribution which leads to traffic skewness $\phi$.
The expected demand completion times of the systems \system, \esn and \rsn 
are as follows. 

\noindent \textbf{For \system} (upper bound):
 \begin{align}\label{eq:dctCerberus}
\dct(\system, \uni(x), k) \le x \cdot \alpha
 \end{align}
 where 
 %$\alpha = \frac{\uni(1,\lflow)}{|\lflow|} \cdot \frac{\crec+\frac{|\lflow|}{r}}{k^*_c}$ 
 $\alpha = \frac{\uni(x,\lflow)}{\cnum^*}\left(\Expct \left[\frac{\crec}{\card{f}}\right]+ \frac{1}{r} \right)$.
 The expected size (of the reciprocal flow sizes) is taken only over large flows. 
 $\cnum^*$ is the optimal number of
 \cs switches, computed from the ratio between the optimal number of  \cs switches to \rs switches denoted by $\rnum^*$:
%  $$
%   \frac{\cnum^*}{\rnum^*} =\frac{\uni(x,\lflow)}{\card{\lflow}}\cdot \frac{\card{\mflow}}{\uni(x,\mflow)(2-\phi_m)} \cdot \frac{\crec+\frac{|\lflow|}{r}}{\rrec+\delta}
%   $$
\begin{align}\label{eq:split}
 \frac{\cnum^*}{\rnum^*} &= \frac{\uni(x,\lflow)}{\uni(x,\mflow)/\card{\mflow}}  \cdot \frac{\Expct \left[\frac{\crec}{\card{f}}\right]+ \frac{1}{r} }{(2-\phi_\mflow)(\rrec+\delta)}
%   \frac{\cnum^*}{\rnum^*} &= \frac{\uni(x,\lflow)/\card{\lflow}}{\uni(x,\mflow)/\card{\mflow}}  \cdot \frac{\crec+\frac{|\lflow|}{r}}{(2-\phi_\mflow)(\rrec+\delta)}
 \end{align} 
where $\phi_\mflow$ is the traffic skewness of the medium size flows.

The threshold for large flows, $\card{\lflow}$ can be computed by: %\chen{add note}
\begin{align}\label{eq:threshold}
  \card{\lflow} \ge \frac{ \crec\cdot|\mflow|\cdot r}{(2-\phi)\cdot r\cdot(\rrec+\delta)-|\mflow|}
%   |f| &\leq \frac{ 15\cdot10^7}{1.1\cdot(2-\phi_t)-1}
\end{align} 

\noindent \textbf{For \rsn} (lower bound):
 \begin{align}\label{eq:dctRotor}
\dct(\rsn, \uni(x), k) &\ge  x \cdot \beta
 \end{align} 
 where $\beta=(2- \phi)\frac{\rrec+\delta}{\delta} $. \\

\noindent \textbf{For \esn} (lower bound):
\begin{align}\label{eq:dctExpander}
\dct(\esn, \uni(x), k) &\ge  x \cdot \gamma
\end{align}
where $\gamma= \EPL(G(k))$ is the expected path length of the $k$-regular expander with $n$ nodes.
\end{theorem}

As we show in the next section, Theorem \ref{thm:linear} enables us to numerically estimate the demand completion times of the different systems given the input parameters (i.e., reconfiguration times, rates, etc.).    
Following Theorem \ref{thm:linear}, a first main observation about \system is that
 the optimal number of switches 
required of types \rs and \cs in \system,
%should be roughly 
is proportional to the share of the different
types of traffic, i.e., of medium and large flows
(Theorem \ref{thm:linear}, Eq. \eqref{eq:split}).

Regarding the skewed traffic, a similar result, but with slightly more complex analysis, 
can be obtained.

\begin{theorem}[Skewed Traffic]\label{thm:sked}
Consider the skewed traffic model, $\skd(x)$ with 
a flow size distribution which leads to traffic skewness $\phi$.
The expected throughput $L$ of the active ToRs for the systems \system, \esn and \rsn can be computed by solving the 
following equations for L:

\noindent \textbf{For \system}:  
\begin{align}
\dct(\system, \skd_L(x), k) &= 1 \;\;\; \Rightarrow \;\;\;
\notag \\
x L \frac{\uni(1,m) +x^*\uni(1,\lflow)}{\card{\mflow}}(2-\phi x)\frac{\rrec+\delta}{\rnum^*} &= 1
%\dct(\system, \skd(x), k) = x(2-\phi x)\alpha'
 \end{align}
where $x^*$ is a function of $L$, denoting the expected fraction of large flows that are routed via \rs switches.
% where $\alpha' = \frac{\uni(1,m) + \overline{\ell^*}}{\card{\mflow}} \frac{\rrec+\delta}{\rnum^*}$.  $\rnum^*$ is the optimal number of \rs switches, $\overline{\ell^*}$ is the number of large flows that need to be sent to the \rs switches, and $\phi$ is the skewness of the traffic that is sent to the \cs switches.
 
\noindent \textbf{For \rsn}:  
\begin{align}
\dct(\rsn, \skd_L(x), k) &= 1 \;\;\;\; \Rightarrow \notag \\ 
%\notag
 %where $\beta'= \frac{\rrec+\delta}{\delta}$, and $\phi$ is the skewness of the traffic.
%\end{align}
%which gives:
%\begin{align}
  L &= \min \left(\frac{1}{x(2- \phi x) \cdot \frac{\rrec+\delta}{\delta}}, 1 \right) 
%\dct(\rsn, \skd(x), k) &=  x(2- \phi x)\beta'
\end{align}

\noindent \textbf{For \esn}: 
\begin{align}
\dct(\esn, \skd_L(x), k) &= 1 \;\;\;\; \Rightarrow \notag \\
%\notag
%\end{align}
%which gives:
%\begin{align}
L &= \min\left(\frac{1}{x \cdot \EPL(G(k)},1\right)
%\dct(\esn, \skd(x), k) &=  x \cdot \gamma
\end{align}
where $\EPL(G(k))$ is the expected path length of a random $k$-regular expander with $n$ nodes.
\end{theorem}

As we show in the next section, Theorem \ref{thm:sked} enables us to numerically estimate the throughput of the different systems given the input parameters (i.e., $x$, reconfiguration times, rates, etc.).    
Following Theorem \ref{thm:sked}, we can state:
 All three systems: \esn, \rsn and \system are throughput-proportional.

\section{Detailed Flow-Level Analysis}\label{sec:flowanalysis}
In this section we present more in-depth analytical results for the performance
of \system. We analyze the demand (aka matrix) completion time:
the \emph{total} time it
 takes to serve a demand matrix (built
from flows) that arrives at time zero.  
We will consider both the \emph{Uniform} and \emph{Skewed} traffic models.

\subsection{Analysis of Uniform Traffic}

Recall that $\tau$ denotes the type of a flow where 
$\tau \in \{\sflow, \mflow, \lflow\}$
is the flow type and let $\card{\mflow}$ and $\card{\lflow}$ denote the \emph{thresholds} sizes to decide the flows type. 
Namely, \emph{small} flows are of size less than $\card{\mflow}$, \emph{medium} flows are of size more than  $\card{\mflow}$ and less than $\card{\lflow}$ and, \emph{large} flows are of size larger than $\card{\lflow}$.
For traffic generated by $\uni(x)$, let $\uni(x, \tau)$
denote the expected number of bytes per second in flows of type $\tau$ when a ToR is working at x\%, namely each of its $k$ links is sending traffic at  rate $r=10 x Gb/s$.
%$\#{\tau}$ denote the number of flows of type $\tau$ when a ToR is working at a 100\% of its capacity, namely each of its $k$ links is sending traffic at a rate of $r=10Gb/s$. 
%Let $\uni(x, \tau)$ denote the flows  of size $\tau$ generated by $\uni(x).$
We set the size of the medium threshold to be exactly $|\mflow| 
= \delta r = 1 Mb$ the flow size that can be transmitted in one slot.
%The time needed to transmit a large flow $\lflow$ on a \cs switch is the $\rrec$ reconfiguration time plus the transmission time $\rrec+\frac{r}{|\lflow|}$.

Consider $k$ switches all of them of the same type $\omega \in \{\s, \r, \c\}$. We denote by  $\dct_{\omega}(D, k)$  the \emph{demand completion time} to serve the demand $D$ using these $k$ switches.
Further, let $\dct(\sys, D, k)$ denote the demand completion time of system $\sys\in\{\esn,\rsn,\system\}$ for demand $D$ using $k$ switches. 
Note that for \esn and \rsn we have $\dct(\esn, D, k)= \dct_{\s}(D, k)$ and 
 $\dct(\rsn, D, k)= \dct_{\r}(D, k)$ respectively, since these systems are built from a single switch type. But this is not the case for \system which is built form a combination of all three switches types. 
 We present the analysis from the easy to the hard case, first \esn, then \rsn and finally \system.

\subsubsection{Analysis of Expander-Net}

We start by (optimistically) approximating the demand completion time of \esn. 
We assume that traffic is distributed along all shortest paths with no delay due to packet loss or congestion.
Hence, the only ``cost'' we consider is related to the path length, that is, each flow consumes bandwidth capacity proportional to the route length (i.e. ``bandwidth tax"). 
For example, if the route length of all flows is two and all ToRs are working uniformly, the maximum achievable load is 50\%; otherwise the total traffic would exceed the network capacity: the number of ToRs times the number of switches times the rate, i.e.,  $n \cdot k \cdot r$. 
Now let $G(\snum)$ denote a (random) $\snum$-regular expander built from $\snum$ (random) matchings and 
let $\EPL(G(\snum))$ denote the \emph{expected path length} of $G(\snum)$. The demand completion time of traffic $\uni(x)$ (per ToR) can be bounded as follows:
 \begin{align}
 \dct_{\s}(\uni(x), \snum) \ge \uni(x) \cdot \frac{\EPL(G(\snum))}{\snum \cdot r} 
 \end{align} 

From this we can compute the bound for an expander made from  $k$ switches and for our traffic model with load $x$: 
 \begin{align}
\dct(\esn, \uni(x), k) &= \dct_{\s}(\uni(x), k)  \notag \\
%&=\dct_{\s}(\uni(x,\sflow) \cup \uni(x,\mflow) \cup \uni(x,\lflow), k) \notag \\
%&=\dct_{\s}(x \cdot (\#\sflow \cdot  |\sflow| + \#\mflow \cdot  |\mflow| + \#\lflow \cdot |\lflow|), k) \notag \\
%&= \frac{x \cdot  (\#\sflow \cdot |\sflow| + \#\mflow \cdot |\mflow| + \#\lflow  \cdot |\lflow|) \cdot \EPL(G(k)) } {k r}  \notag \\
&= \frac{x\cdot  k\cdot  r \cdot \EPL(G(k)) }{ r\cdot k  } \\ \notag 
&
= x \cdot \EPL(G(k))
 \end{align}
%\khen{not sure the reader (or me) will understand why we $x*k*r$ here }
Therefore the completion time is linear in $x$. Using our test parameters from Table \ref{tab:vars} we have found that a (random) 32-regular expander with 256 nodes has an expected path length that is about 1.85. 
%\khen{should we mention how we got this number?} \chen{good point, not sure...}
%\khen{we need to rethink this formula perhaps why dived by k?? }

\subsubsection{Analysis of Rotor-Net}
Next we consider \rsn.
First we consider a completely uniform demand between all possible pairs denoted as $\overline{\uni}$
(arriving at time zero). In this all-to-all case, \rsn will be almost optimal
by serving requests in each slot according to the current matching 
of each switch; all ports will be continuously %working
operating at 100\% throughput, sending flows directly (in a single hop) from source to destination. The only inefficiency will be due to the reconfiguration time (the ``latency tax''),
to reconfigure between slots. In our setting this 
overhead is equal to $\rrec/(\rrec + \delta)$, about 9\% if we use Table \ref{tab:vars} parameters. 
So the demand completion time of $k$ \rs switches and such uniform demand $\overline{\uni}$ is for a single ToR (and for the system):
(number of slots in $\overline{\uni}$) $/$ (number switches) $\times$ (time for a slot). Formally,

 \begin{align}
 \dct_{\r}(\overline{\uni},k) =\frac{\overline{\uni}}{k \cdot r} \cdot \frac{\rrec+\delta}{\delta} = \frac{\overline{\uni}}{\delta \cdot r} \cdot \frac{\rrec+\delta}{k}=  \frac{\overline{\uni}}{\card{\mflow}} \cdot \frac{\rrec+\delta}{k}
 \end{align} 
 
 Next we consider the case of traffic $\uni(x)$, in this case all ToRs sample flow sizes from the same flow distribution $\D$, but flows can have different sizes so we cannot assume the traffic is uniform among all pairs or all-to-all.
 %\khen{so now we have  $\overline{\uni}$, $\D$ D, U(x) that describe very similar things...}
Dealing with non-uniform flows is more complex, since the number of larger flows could be  relativity small, potentially leaving many links in each slot inactive. 

%start copy%
A \rsn overcomes this problem by using Valiant routing (load balancing) \cite{rotornet} 
where flows and packets can be sent via two hops and not directly. 
Flows (or packets) that takes two hops may take additional capacity from the network (i.e. ``bandwidth tax"). 
%Some flows are therefore sent via two hops or one hop, this depends on the uniformity or sparsity of the demand.
We model this situation with a \emph{traffic skewness} parameter $0 \le \phi \le 1$ which approximates the fraction of bytes that a ToR in \rsn sends using one hop. 
When traffic is close to uniform among destinations for a ToR, then $\phi$ will be close to 1 since there are no available other ToRs to help the current ToR. When the destination for a ToR are skewed. e.g., one large flow toward a single destination, $\phi$ will be close to 0 and most of the traffic will be sent via two hops, taking advantage of destinations that are free to help.
To approximate $\phi$ for a give distribution (or empirical distribution) we define it as one minus the variation distance \cite{rosenthal1995convergence} from the uniform distribution. For a finite state PDF $P=\{p_1, p_2, \dots ,p_n\}$ the variation distance  from the uniform distribution is define as:
\begin{align}
    \Delta(P) = \frac{1}{2}\sum_{i=1}^n \card{p_i - \frac{1}{n}}
\end{align}

$\Delta(P)$ measures the probably mass above the average.
If, for a ToR, $P$ represents the distribution of the destinations' load, $\Delta(P)$ will capture the fraction of packets that can benefit from sending by two hops (using the help of destinations that have load bellow the average) and $\phi =1 - \Delta(P)$ is the fraction of messages that will use a single hop.
Therefore the average number of hops that a packet takes is: $1 \cdot \phi + 2(1-\phi)=2-\phi$.  Since ToRs are symmetric in this traffic model (i.e., $\uni(x)$) a ToR that sends $(1-\phi)$ of its traffic via two hops, asking the help of other ToRs, would expect similar requests from him to help other ToRs. 
This means that the expected total traffic need to be sent by a ToR will be $\uni(x)(2-\phi)$
and, we can optimistically assume
%\khen{why optimistically?} 
it will now be divided uniformly among destinations (otherwise the DCT will be only larger) and 
we can formally generalize the DCT lower bound to:
  \begin{align}\label{eq:krotor}
 \dct_{\r}(\uni(x),k) =\frac{\uni(x)}{\card{\mflow}}(2- \phi)\frac{\rrec+\delta}{k}
 =\frac{\uni(x)}{kr}(2- \phi) \frac{\rrec+\delta}{\delta}
 \end{align} 

Since \rsn is composed of $k$ \rs switches we can now bound the demand completion time of load $x$ for \rsn by:

 \begin{align}\label{eq:largerotor}
 \dct(\rsn, \uni(x), k) &= \dct_{\r}(\uni(x), k) \notag \notag\\
&= \frac{x r k}{\card{\mflow}}(2- \phi) \frac{\rrec+\delta}{k} \notag\\
&=x(2- \phi)\frac{\rrec+\delta}{\delta}
 \end{align} 

If $\phi$ is a constant then this is a linear function.
In practice $\phi$ can vary, but nevertheless the function can still be approximated well by a linear function.

\subsubsection{Analysis of \system}

We now turn to discuss \system whose analysis
is a bit more complex. 
Let's start with large flows which, 
according to Algorithm \ref{alg:cachenet}, 
 are transmitted via the demand-aware switches. The cache component operates by reconfiguring a direct link form the source of a flow to its
destination. The flow is then transmitted along a single hop. Assuming the reconfiguration time of a single cache switch is $\crec$ and the transmission time of a single large flow $f$ of size $|f|$ is $|f|/r$,
the flow completion time for a single flow on a single switch is   $\crec +\frac{|f|}{r}$.   

\para{Finding the threshold for large flows.} To determine the threshold we would like to find a value for a flow size $|f|$ for which sending the flow $f$ on a $k$-rotor switch network is slower than transmitting the same flow on a $k$-cache switch network. Following Eq. \eqref{eq:krotor} we note that the transmission time is a function of the global traffic skewness $\phi$:
\begin{align}
 %\dct_{\c}(f, k) &\leq  \dct_{\r}(f, k) \Rightarrow \notag \\
   \frac{\crec+\frac{|f|}{ r}}{k} &\leq \frac{|f|}{|\mflow|}\cdot(2-\phi)\cdot \frac{\crec+\delta}{k}  \notag \\
  \crec &\leq \frac{|f|}{|\mflow|}\cdot (2-\phi) \cdot (\rrec+\delta)-\frac{|f|}{r} \notag\\
  \frac{ \crec\cdot\mflow\cdot r}{(2-\phi)\cdot r\cdot(\rrec+\delta)-|\mflow|} &\leq \card{f}
%   |f| &\leq \frac{ 15\cdot10^7}{1.1\cdot(2-\phi_t)-1}
 \end{align} 

When we use the parameters of Table \ref{tab:vars} we have that 
for $\phi=0$ (all packets of the flow are sent via two hops in \rsn), 
$\card{f} = 15MB$, and for $\phi=1$ (all packets of the flow are sent via one hops in \rsn), $\card{f}=187.5MB$. The threshold we use in our evaluation is $125MB$.

\para{The demand completion time.}
For a given partition of the $k$ switches to the three types of switches: $\snum$ \ss switches, $\rnum$ \rs switches and $\cnum$ demand-aware switches, and a uniform traffic model, $\uni(x)$ with load $x$, the demand completion time of \system is the maximal completion time among the three sub-components, formally,

\begin{align}\label{eq:max}
\dct(\system, \uni(x), k) = \max
\begin{cases}
\dct_{\s}(\uni(x,\sflow), \snum)\\%=\dct_{\s}(x \cdot \#\sflow  |\sflow|, \snum)\\
\dct_{\r}(\uni(x,\mflow), \rnum)\\%=\dct_{\r}(x \cdot \#\mflow  |\mflow|, \rnum)\\
\dct_{\c}(\uni(x,\lflow), \cnum)%=\dct_{\c}(x \cdot \#\lflow  |\lflow|, \cnum)
\end{cases}
 \end{align}

Assuming the sources and destinations are distributed uniformly and there are $\cnum$ demand-aware switches,
then the demand completion time of a single ToR (and by symmetry all ToRs) is approximated by

 \begin{align}
 \dct_{\c}(\uni(x,\lflow), \cnum) =\frac{1}{\cnum} \sum_{f \in \uni(x,\lflow)} (\crec+\frac{|f|}{r}) 
 \end{align} 
 
In the worst case we have $\uni(x,\lflow)/\card{\lflow}$ large flows, each of size $|\lflow|$: as the flows get larger, the reconfiguration time $\crec$ is better amortized by the transmission time of the flow. The upper bound of the demand completion time is then

\begin{align}\label{eq:fctcache}
 \dct_{\c}(\uni(x,\lflow), \cnum) \le \frac{\uni(x,\lflow)}{ |\lflow| }\cdot \frac{\crec+\frac{|\lflow|}{r}}{\cnum}
 = \frac{\uni(x,\lflow)}{\cnum}\left(\frac{\crec}{\card{\lflow}}+ \frac{1}{r}\right)
 \end{align}

When the threshold $\card{\lflow}$ is far from the average large flow it is better to take the expected completion time.

\begin{align}\label{eq:fctcache}
 \dct_{\c}(\uni(x,\lflow), \cnum) &= \sum_{f} \left(\frac{\Pr (f)\uni(x,\lflow)}{|f|} \frac{\crec+\frac{|f|}{r}}{\cnum}\right) \\
%\notag \\ &
&= \sum_{f} \left(\frac{\Pr (f)\uni(x,\lflow)}{\cnum}\left(\frac{\crec}{\card{f}}+ \frac{1}{r}\right) \right) 
\notag \\ & \frac{\uni(x,\lflow)}{\cnum}\left(\Expct \left[\frac{\crec}{\card{f}}\right]+ \frac{1}{r} \right)
 \end{align}

Next we discuss how to find and optimal partition of the switches.
For example, assuming $\snum = 5$ and that the completion time for
the expander component is negligible (due to low traffic volume), 
the optimal division of the switches, denoted as $\rnum^*$ and $\cnum^*$,
is such that the 
completion time of the corresponding components will be identical. This follows from the fact that the completion 
time of each sub-component, as shown above, is monotonically decreasing in the number of switches. Equalizing the two components allows us to compute the optimal 
number of switches, as follows (where $\phi_\mflow$ denote the skewness of the medium size traffic):

\begin{align}
 \dct_{\c}(\uni(x,\lflow) , \cnum^*) &=  \dct_{\r}(\uni(x,\mflow), \rnum^*) \Rightarrow \\
%  \frac{\uni(x,\lflow)}{|\lflow|}\cdot \frac{\crec+\frac{|\lflow|}{r}}{\cnum^*} 
  \frac{\uni(x,\lflow)}{\cnum^*}\left(\Expct \left[\frac{\crec}{\card{f}}\right]+ \frac{1}{r} \right) &= 
  \frac{\uni(x,m)}{\card{\mflow}}\cdot (2- \phi_\mflow) \cdot \frac{\rrec+\delta}{\rnum^*}  \Rightarrow \notag \\
   \frac{\cnum^*}{\rnum^*} &= \frac{\uni(x,\lflow)}{\uni(x,\mflow)/\card{\mflow}}  \cdot \frac{\Expct \left[\frac{\crec}{\card{f}}\right]+ \frac{1}{r} }{(2-\phi_\mflow)(\rrec+\delta)} \notag
% \frac{\cnum^*}{\rnum^*} &= \frac{\uni(x,\lflow)}{\uni(x,\mflow)} \cdot \frac{\rrec+\frac{|\lflow|}{r}}{(2-\phi_\mflow)\cdot(\rrec+\delta)}
 \end{align} 
For example, for a case study distribution and traffic $\uni(0.5)$ we calculated $\phi_\mflow$ and by plugging in the values from Table \ref{tab:vars},
we get that $\cnum^* =16$ and $\rnum^*=16$ (using rounding
and recalling that $\rnum^* + \cnum^* = 32$). 

Following Eq.~\eqref{eq:fctcache} we can now approximate the demand completion time of \system as the completion time of the $\cnum^*$ cache switches (recall that it is equal to the completion time of the $\rnum^*$ switches).

\begin{align}
\dct(\system, \uni(x), k) \le x \cdot \frac{\uni(x,\lflow)}{\cnum^*}\left(\Expct \left[\frac{\crec}{\card{f}}\right]+ \frac{1}{r} \right)
% \frac{\uni(1,\lflow)}{|\lflow|} \cdot \frac{\crec+\frac{|\lflow|}{r}}{k^*_c}
\end{align}

\subsection{Analysis of Skewed Traffic}\label{sec:analysisskewed}

Next we analyze the skewed traffic model %which is more skewed, % a more skewed traffic, 
$\skd(x)$, where only a fraction $x$ of the ToRs 
are active and operating at maximum rate.
Let $\skd_L(x)$ denote the case where each of the active ToRs works at a load $L$ and $\skd_1(x)=\skd(x)$.
Our goal in this scenario is to design a network which allows for a maximal possible per-ToR throughput $L$.
Let $L^*(x) = \argmax_L \skd_L(x) \le 1$ and $L^* \le 1$ is highest throughput that the network can support (given $x$).
%, denoted as $Th(x) \le 1$,
%where a throughput of one means that each port can send with maximal rate $r$.

The assumption of skewed traffic is more realistic than
uniform traffic, but it is also harder to analyze. 
The main challenge arises from the fact that 
under the skewed traffic model, only a fraction $x$ of the ToRs are active, but in principle, it may be possible to utilize the $1-x$ fraction of non-active ToRs
to help the active ones. The question is if this can be done, and what is the maximum achievable throughput $L^*(x)$. 
%The question we ask is:
%what is the maximum per-ToR throughput $L$ that the network
%can support, for a given network load $x$? 
We formally analyze this scenario by computing the demand completion time for $\skd(x)$ (with ports working at full throughput $r$). If the completion time is less than one second,  
we say that the throughput is one; otherwise we adjust the throughput, $L$ to find the largest throughput for which the completion time is less than one second.
Let $\skd(x, \tau)$, as before,
denote the expected number of bytes per second in flows of type $\tau$ generated by $\skd(x)$, similarly we have $\skd_L(x, \tau)$.

We start again with the \esn. We optimistically assume the \esn has all the good properties expanders  should have \cite{hoory2006expander},
i.e.,  large expansion, multiple disjoint paths, small mixing times, etc. 
Basically these properties guarantee that even
for a small set of communicating ToRs, the traffic 
will efficiently spread across the entire network, utilizing the
non-active ToRs as much as possible. Therefore, as before, 
we assume only the capacity restriction and consider the demand completion
time as:

\begin{align}\label{eq:xpander2}
%&
\dct(\esn, \skd_L(x), k) = L x \cdot \EPL(G(k))
\end{align}
We can find the throughput by solving for $L$.  
\begin{align}\label{eq:xpander3}
L x \cdot \EPL(G(k)) = 1
\end{align}
From this we can find $L^*(x)$ as: 
\begin{align}\label{eq:xpander}
L^*(x) %&= \min\left(\frac{1}{\dct(\esn, \skd(x), k)}, 1\right) \notag \\
&= \min\left(\frac{1}{x \cdot \EPL(G(k)},1\right)
 \end{align}

Following the definitions and observation in~\cite{xpander} 
%Figure
%and in particular Figure~2 within,
%End Figure
the above Eq.~\eqref{eq:xpander} shows 
that a static expander will be a \emph{throughput-proportional network}, namely it will be ``able to distribute its capacity evenly across only the set of servers with traffic demands''~\cite{xpander}.

What about a rotor-based network or a network based on \system? 
Achieving throughput-proportionality seems to be non-trivial.
For example if the load is 50\%, how can the network exploit
the capacity of the 50\% non-active ToRs to help the active ones 
work at 100\% throughput? In~\cite{xpander},
it was shown that the fat-tree is not throughput-proportional.

Interestingly, we can show that \rs switches actually work very well
in this scenario and that \rsn is also a throughput proportional 
network. 
This is a novel result and was not discussed in previous work \cite{xpander,rotornet,opera}. 
The result is due to the Valiant routing property of \rsn: if every flow is transmitted
using this mechanism, flows will be forwarded using the non-active ToR, 
keeping \emph{all} switch ports sending at full rate. Moreover, 
if a fraction $x$ of the network is active and generates
a uniform traffic (within the active fraction), this means that 
a fraction $x$ of the \emph{time}, switch ports can 
directly communicate to their destinations, and only
a $1-x$ fraction of the time flows will need to use two hops. 
If the traffic within the $x$ fraction of active ToRs is non-uniform we can again use the skewness parameter $\phi$ to approximate the fraction of traffic that needs one hop and the fraction that needs two hops. The average number of hops will then be $x(\phi + 2(1-\phi)) + 2(1-x)=2-\phi x$. 
The total amount of traffic to be sent 
in the whole network will now be $\skd(x) (2-\phi x) = x n k r (2-\phi x)$, but since we are using two hops, it will be uniformly divided among ToRs and destinations, and
we can approximate the demand completion time of a single ToR as:
%\khen{ check and rethink this?}
%So for medium (and small) flows, we can approximate the demand completion time:
%with some abuse of notation let $|\mflow|(x)$ denote the case were only x\% of the nodes generate the medium flows (but at full rate).
%Then:
%\stefan{double check it:}
\begin{align}\label{eq:rotskwedmedium} 
 \dct(\rsn, \skd(x), k) &= \dct_{\r}(\skd(x), k)
\notag \\
 &
= \frac{\skd(x)}{\card{\mflow}}(2- \phi x) \cdot \frac{\rrec+\delta}{n k}
\notag \\
&
= x(2- \phi x) \cdot \frac{\rrec+\delta}{\delta}
 \end{align} 

And find the throughput by solving for $L$.
\begin{align}\label{eq:rotskwed} 
 \dct(\rsn, \skd_L(x), k) &= \dct_{\r}(\skd_L(x), k) = 1 \Rightarrow \notag \\
 1 &= L x(2- \phi x) \cdot \frac{\rrec+\delta}{\delta}
 \end{align} 
 
and 
\begin{align}\label{eq:rotskwedthr} 
 L^*(x) = \min \left(\frac{1}{x(2- \phi x) \cdot \frac{\rrec+\delta}{\delta}}, 1 \right)
 \end{align}
 
The above equation implies that \rsn is throughput-proportional. For the setting of Table \ref{tab:vars} and our Case Study distribution with $\phi = 0.49$ it supports up to 50\% of the ToRs working at full rate, the same as for the expander.

\begin{observation}\label{obs:throughput}
\rsn is a throughput-proportional network.
\end{observation}
 
We next discuss \system. We already know that the \rs switches are throughput-proportional, so what about the demand-aware switches? 
Due to reconfiguration time, the demand-aware switches may not be able to
serve all large flows generated by the active ToRs. Furthermore,
our current design of demand-aware switches does not support 2-hops routing.
The ``latency tax'' due to reconfigurations in our numerical example
is at most 15\%, $\frac{\crec}{|\lflow|/r}$, so when an active ToR is working at a full rate, it cannot send all of its large flows to the demand-aware switches.

Let $z$ denote the expected fraction of large flows that $\cnum$ demand-aware switches can send in a second for a given ToR. We can find it using Eq. \eqref{eq:fctcache}

 \begin{align}\label{eq:xstar}
 \dct_{\c}(\uni(z,\lflow), \cnum) 
 &= \frac{\uni(z,\lflow)}{\cnum}\left(\Expct \left[\frac{\crec}{\card{f}}\right]+ \frac{1}{r} \right) = 1 \Rightarrow \notag\\
 1 &= z \frac{\uni(1,\lflow)}{\cnum}\left(\Expct \left[\frac{\crec}{\card{f}}\right]+ \frac{1}{r} \right)  \Rightarrow \notag \\
 z &= \frac{\cnum}{\uni(1,\lflow)\left(\Expct \left[\frac{\crec}{\card{f}}\right]+ \frac{1}{r} \right)}
 \end{align} 

Using $z$ and following Algorithm~\ref{alg:cachenet},
we can approximate the expected fraction of large flows, denoted as $x^*$ per ToR which cannot fit the cache and which will be sent to the rotor switches when working at rate $L$ (in $\skd_L(x)$), by:
\begin{align}
x^* = \max (\frac{L  -  z}{L}, 0)
\end{align}
The amount of such traffic per ToR will be $x^*L\uni(1,\lflow) = x^*\uni(L,\lflow)$.
Assuming that small flows are transmitted via static switches in \system 
with negligible completion time, we can compute the demand completion time of \system as the demand completion time of the $\rnum^*$~ \rs switches
(since the \cs switches are set up to have demand such that they finish at $1 s$).
Following Eq.~\eqref{eq:rotskwedmedium}, we have:
 
\begin{align}\label{eq:sysskewd}
&\dct(\system, \skd(x), k) = \notag\\
&=\dct_{\r}(\skd(x,m) \cup  xn x^*\uni(1,\lflow), \rnum^*) \notag\\
&=  \frac{\skd(x,m) + xnx^*\uni(1,\lflow)}{\card{\mflow}}(2-\phi x)\frac{\rrec+\delta}{n \rnum^*} \notag \\
&=  x \frac{\uni(1,m) +x^*\uni(1,\lflow)}{\card{\mflow}}(2-\phi x)\frac{\rrec+\delta}{\rnum^*}
 \end{align} 
 
%   \begin{align}\label{eq:sysskewd}
% \dct(\system, \skd(x), k) &= \dct_{\r}(\skd(x,m) \cup  x n \overline{\ell^*}, \rnum^*) \\
% &=  \frac{\skd(x,m) + x n\overline{\ell^*}}{\card{\mflow}}(2-\phi x)\frac{\rrec+\delta}{n \rnum^*} \notag \\
% &=  x \frac{\uni(1,m) + \overline{\ell^*}}{\card{\mflow}}(2-\phi x)\frac{\rrec+\delta}{\rnum^*}
% \notag
%  \end{align}
 
%\stefan{add an observation on throughput-proportionality?}

%\begin{align}
%\dct(\system, \skd(x), k) =  x\left ((2-x)\#\mflow  |\mflow|  + 2 \overline{\ell^*}  |\lflow| \right )  \frac{\rrec+\delta}{|\mflow| k}
% \end{align}
\noindent
where $\phi$ is the skewness parameter that fit the traffic generated by $\skd(x,m) \cup  x^*\uni(1,\lflow)$. 

To find $L^*(x)$ we again need to solve for $L$ s.t.

\begin{align}\label{eq:sysskewdthr}
1 &=\dct(\system, \skd_L(x), k) \Rightarrow \notag \\
1 &=  x \frac{\uni(L,m) +x^*\uni(L,\lflow)}{\card{\mflow}}(2-\phi x)\frac{\rrec+\delta}{\rnum^*} \Rightarrow \notag \\
1 &=  x L \frac{\uni(1,m) +x^*\uni(1,\lflow)}{\card{\mflow}}(2-\phi x)\frac{\rrec+\delta}{\rnum^*} 
 \end{align} 

and $L^*(x)$ can be found numerically.
%\begin{align}\label{eq:cachethr} 
%  L^*(x) = \min \left(\frac{\card{\mflow} \rnum^*}{x(\uni(1,m) +x^*\uni(1,\lflow))(2-\phi x)}, 1 \right)
%  \end{align}

%Figure
% Figure \ref{fig:Thsystems} shows the throughput of \esn, \rsn and \system following Eq.~\eqref{eq:xpander}, \eqref{eq:rotskwedmedium} and \eqref{eq:sysskewd}, both for the Case Study and for the Datamining.

% All systems are throughput-proportional with \system preforming better
% at higher loads (> 70\%).
% further by making these links static and random, and support the expander links. 

% We note that Figures \ref{fig:DCTsystems} and \ref{fig:Thsystems} 
% are related: for each system, the maximum supported throughput $L$ 
% under maximum load, i.e.,  $\dct(\skd_L(1)) = 1$ in Figure~ \ref{fig:Thsystems}
% exactly corresponds to the point on Figure \ref{fig:DCTsystems} where the load is $L$ and the demand completion time is $1 s$, i.e., $\dct(U(L))=1$.
%End Figure
%Thus 
%\khen{need info from fig?}

We can derive the following theorem.
\begin{theorem}[Skewed Traffic]\label{thm:skedAppendix}
Consider the skewed traffic model, $\skd(x)$ with 
a flow size distribution which leads to traffic skewness $\phi$.
The expected throughput $L$ of the active ToR for the systems \system, \esn and \rsn can be computed by solving the 
following equations for L:

\noindent \textbf{For \system}:  
\begin{align}
\dct(\system, \skd_L(x), k) &= 1 \Rightarrow \notag \\
x L \frac{\uni(1,m) +x^*\uni(1,\lflow)}{\card{\mflow}}(2-\phi x)\frac{\rrec+\delta}{\rnum^*} &= 1
%\dct(\system, \skd(x), k) = x(2-\phi x)\alpha'
 \end{align}
% where $\alpha' = \frac{\uni(1,m) + \overline{\ell^*}}{\card{\mflow}} \frac{\rrec+\delta}{\rnum^*}$.  $\rnum^*$ is the optimal number of \rs switches, $\overline{\ell^*}$ is the number of large flows that need to be sent to the \rs switches, and $\phi$ is the skewness of the traffic that is sent to the \cs switches.
 
\noindent \textbf{For \rsn}:  
\begin{align}
\dct(\rsn, \skd_L(x), k) &= 1 \Rightarrow \notag \\
\min \left(\frac{1}{x(2- \phi x) \cdot \frac{\rrec+\delta}{\delta}}, 1 \right) &=  L^*(x) 
%\dct(\rsn, \skd(x), k) &=  x(2- \phi x)\beta'
 \end{align} 
 %where $\beta'= \frac{\rrec+\delta}{\delta}$, and $\phi$ is the skewness of the traffic.

\noindent \textbf{For \esn}: 
\begin{align}
\dct(\esn, \skd_L(x), k) &= 1 \Rightarrow \notag \\
\min\left(\frac{1}{x \cdot \EPL(G(k)},1\right) &= L^*(x)
%\dct(\esn, \skd(x), k) &=  x \cdot \gamma
\end{align}
where $\EPL(G(k))$ is the expected path length of a random $k$-regular expander with $n$ nodes.
\end{theorem}

\begin{table}[t!]%[htp]
%\small
\begin{center}
\begin{tabular}{|c|m{4cm}|c|}
 \hline
Symbol & Def & Default Value \\ \hline  \hline
$n$ & number of ToRs & 256 \\ \hline
$N$ & set of ToRs & $1 \dots n$ \\ \hline
$k$ & number of switches  & 37\footnote{If not mentioned otherwise we assume that in all systems there are 5 static switches to take care of small flows and focus on how to divide the rest of the 32 switches.} \\ \hline
$SW$ & the set of switches & $sw_1, \dots, sw_k$  \\ \hline
$\rrec$ & \rs switch configuration time   & $10 \mu s$ \\ \hline
$\crec$ & \cs switch configuration time  & $15 ms$ \\ \hline
$\delta$ & \rs switch slot time  &$100 \mu s$ \\ \hline
$r$ & NIC transmission rate   & $40Gbps$ \\ \hline \hline
$\snum$ & number of \ss switches  & varies \\ \hline
$\rnum$ & number of \rs switches  & varies \\ \hline
$\cnum$ & number of \cs switches  & varies \\ \hline
$\phi$ &   skewness & varies \\ \hline
\end{tabular}
\end{center}
\caption{Variables and parameters definitions and default values}
\label{tab:vars}
\end{table}%
{\balance
  \bibliographystyle{abbrv} 
   \balance
\bibliography{bibs/literature,bibs/cerberus, bibs/bibly}
}

\end{document}